\documentclass[conference]{IEEEtran}
\IEEEoverridecommandlockouts
\usepackage{amsmath,amssymb,amsfonts}
\usepackage{algorithmic}
\usepackage{graphicx}
\usepackage{textcomp}
\usepackage{xcolor}
\def\BibTeX{{\rm B\kern-.05em{\sc i\kern-.025em b}\kern-.08em
    T\kern-.1667em\lower.7ex\hbox{E}\kern-.125emX}}

\usepackage{gensymb} 
\usepackage{subcaption} 
\usepackage{multirow} 
\usepackage{booktabs} 
\usepackage{array}
\usepackage{xspace}
\usepackage{colortbl}

\usepackage[numbers,sort&compress]{natbib}

\usepackage{url}

\usepackage[left=0.65in,right=0.65in,top=0.75in,bottom=1.1in]{geometry}
\newcolumntype{L}[1]{>{\raggedright\let\newline\\\arraybackslash\hspace{0pt}}m{#1}}
\newcolumntype{C}[1]{>{\centering\let\newline\\\arraybackslash\hspace{0pt}}m{#1}}
\newcolumntype{R}[1]{>{\raggedleft\let\newline\\\arraybackslash\hspace{0pt}}m{#1}}

\newcommand{\ie}{\textit{i.e.,}\xspace}

\newcommand{\att}{MNO-A\xspace}
\newcommand{\tmo}{MNO-B\xspace}
\newcommand{\vzn}{MNO-C\xspace}
\newcommand{\attandtmo}{MNO-A \& B\xspace}

\definecolor{light-gray}{gray}{0.95}

\begin{document}

\title{Indoor Sharing in the Mid-Band: A Performance Study of Neutral-Host, Cellular Macro, and Wi-Fi}

\author{
\IEEEauthorblockN{
Joshua Roy Palathinkal\IEEEauthorrefmark{1},
Muhammad Iqbal Rochman\IEEEauthorrefmark{1},
Vanlin Sathya\IEEEauthorrefmark{2},
Mehmet Yavuz\IEEEauthorrefmark{2},
and Monisha Ghosh\IEEEauthorrefmark{1}}
\IEEEauthorblockA{
\IEEEauthorrefmark{1}Department of Electrical Engineering, University of Notre Dame, 
\IEEEauthorrefmark{2}Celona, Inc.\\ 
Email: \IEEEauthorrefmark{1}\{jpalathi,mrochman,mghosh3\}@nd.edu, \IEEEauthorrefmark{2}\{vanlin,mehmet\}@celona.io}
}

\maketitle

\begin{abstract}

Indoor environments present a significant challenge for wireless connectivity, as immense data demand strains traditional solutions. Public Mobile Network Operators (MNOs), utilizing outdoor macro base stations (BSs), suffer from poor signal penetration. Indoor Wi-Fi networks, on the other hand, may face reliability issues due to spectrum contention. Shared spectrum models, particularly the Citizens Broadband Radio Service (CBRS) utilized by private 4G/5G networks, have emerged as a promising alternative to provide reliable indoor service. Moreover, these private networks are equipped with the neutral-host (NH) model, seamlessly offloading indoor MNOs' traffic to the private CBRS network. This paper presents a comprehensive, in-situ performance evaluation of three co-located technologies utilizing mid-bands spectrum (1--6~GHz)---a CBRS-based NH network, public MNO macro networks, and a Wi-Fi~6 network---within a large, big-box retail store characterized by significant building loss.
Our analysis demonstrates: (i) the NH network provides superior indoor coverage compared to MNO macro, requiring only six CBRS devices (CBSDs)---versus 65 Access Points (APs) for enterprise Wi-Fi---to achieve full coverage, with a median building loss of $26.6$~dB ensuring interference-free coexistence with outdoor federal incumbents; (ii) the NH network achieves substantial indoor throughput gains, with per-channel normalized throughput improvements of $1.44\times$ and $1.62\times$ in downlink (DL), and $4.33\times$ and $13\times$ in uplink (UL), compared to 4G and 5G macro deployments, respectively; (iii) the NH deployment achieves a median indoor aggregated physical (PHY)-layer DL throughput gain of $2.08\times$ over 5G macro deployments indoors, despite utilizing only $40$~MHz of aggregated bandwidth compared to $225$~MHz for 5G macro; and (iv) the NH deployment also outperforms Wi-Fi in application-layer HTTP DL performance by $5.05\times$.
The findings offer critical insights into the practical capabilities of shared spectrum models to inform the potential indoor sharing of newly proposed frequencies, such as the 3.1--3.45 GHz and 7.125--8.4 GHz bands.

\end{abstract}



\section{Introduction}\label{sec:intro}

The U.S. Federal Communications Commission (FCC) has spearheaded efforts to meet escalating data demands by significantly expanding access to mid-band (1--6~GHz) frequencies. This initiative includes the auction of several key bands for 5G services: the 2.5--2.69~GHz Broadband Radio Service, the 3.7--3.98~GHz C-band, and the 3.45--3.55~GHz Department of Defense (DoD) band. However, indoor data demands present unique challenges: up to 80\% of U.S. mobile data originates or terminates indoors~\cite{ericsson2023mobility}, with over 80\% offloaded to Wi-Fi~\cite{calabrese2024frontier}.
Despite Wi-Fi's expansion into 6~GHz, achieving reliable indoor wireless coverage remains challenging due to Wi-Fi's lower power and contention-based Medium Access Control (MAC) layer. While the 4G/5G scheduled MAC is more robust, indoor coverage from outdoor macro deployments is similarly challenged due to building penetration losses. Indoor solutions such as Distributed Antenna Systems (DAS) for cellular tend to be very expensive and are not scalable.

\begin{figure*}
    \begin{subfigure}{.3\textwidth}
    \includegraphics[width=\linewidth]{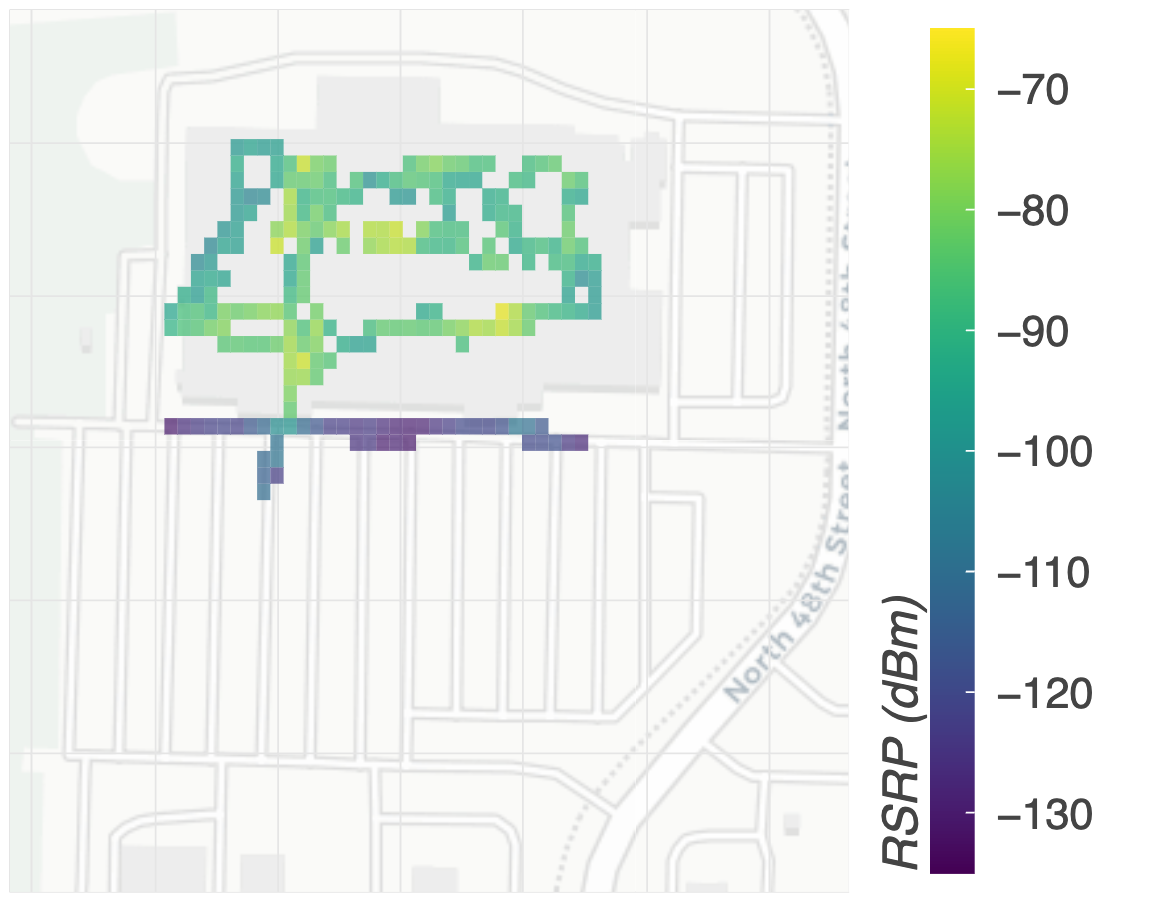}
    \vspace{-1.5em}
    \caption{MNO NH.}
    \label{fig:map_mno_nh}
    \end{subfigure}
    \hfill
    \begin{subfigure}{.3\textwidth}
    \includegraphics[width=\linewidth]{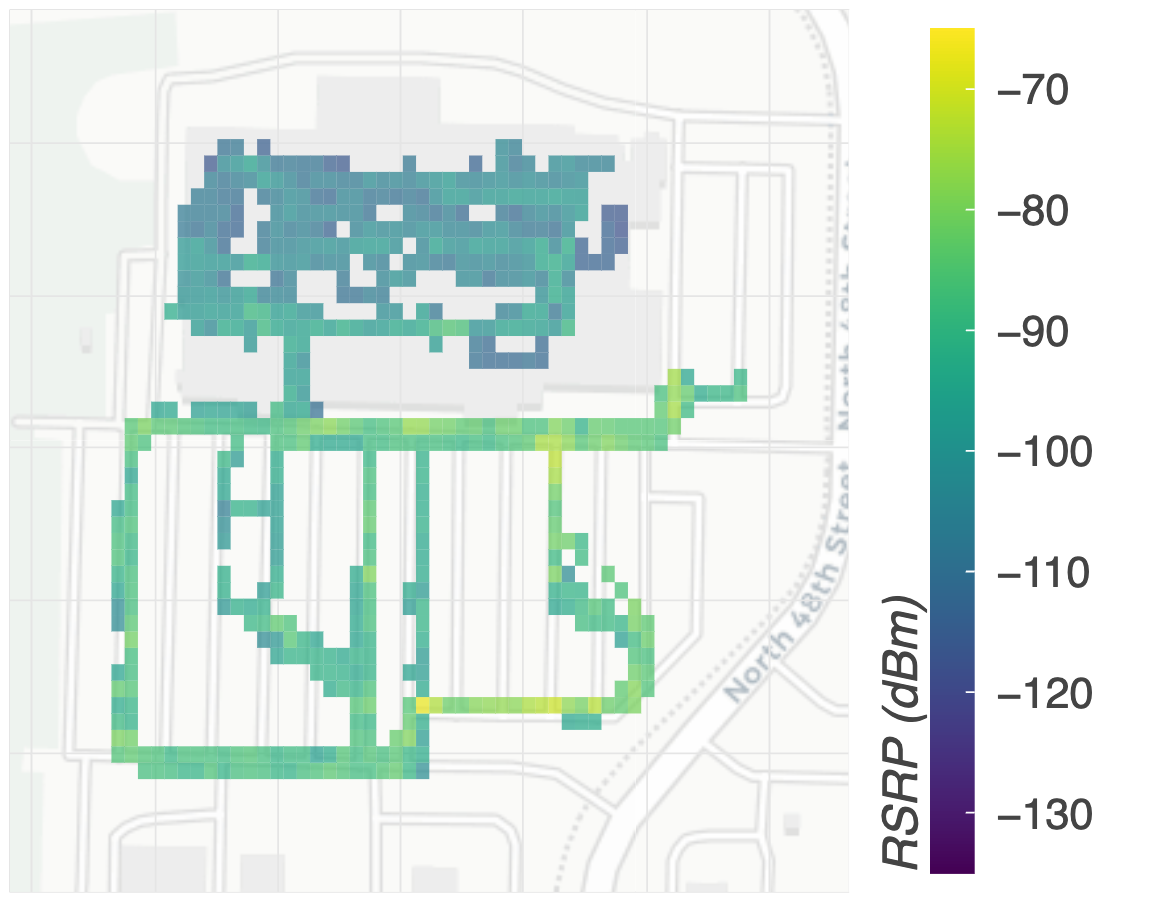}
    \vspace{-1.5em}
    \caption{MNO macro 4G.}
    \label{fig:map_mno_lte}
    \end{subfigure}
    \hfill
    \begin{subfigure}{.3\textwidth}
    \includegraphics[width=\linewidth]{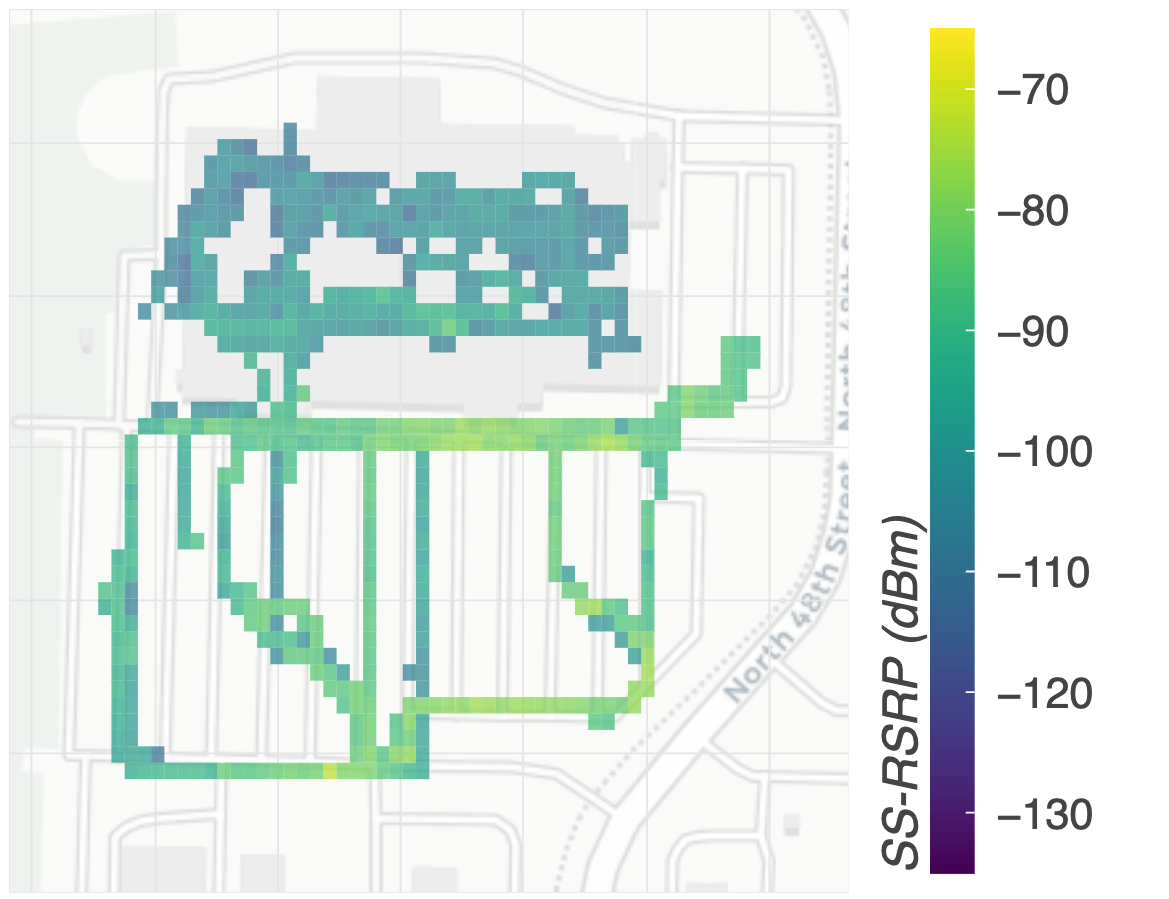}
    \vspace{-1.5em}
    \caption{MNO macro 5G.}
    \label{fig:map_mno_nr}
    \end{subfigure}
    \caption{RSRP heatmap comparison across NH and MNO macro deployments.}
    \label{fig:map_coverage}
    \vspace{-1.5em}
\end{figure*}

To that end, a spectrum sharing model with outdoor incumbents offers an indoor coverage solution, leveraging building loss to separate indoor and outdoor emissions. For instance, the Low Power Indoor (LPI) mode of operation in 6~GHz has shown low probability of harmful interference to outdoor fixed-link incumbents~\cite{dogan2023evaluating,dogan2023indoor}, while the 3.55--3.7~GHz Citizens Broadband Radio Service (CBRS) offers low power indoor operations to enable coexistence with outdoor incumbents like Navy radar~\cite{tusha2025comprehensive}. More recently, the neutral-host (NH) model deployed by private 4G/5G operators in the CBRS band enables seamless offloading of outdoor Mobile Network Operator (MNO) traffic (data and call) to the private indoor radios to enhance connectivity and performance. 
While prior work has compared NH to MNO macro~\cite{bajracharya2022neutral,sathya2023nh} and Wi-Fi~\cite{sathya2023warehouse}, these studies often lack a deep analysis of the underlying performance gains. Our recent study~\cite{rochman2025neutralhosts} analyzed the 4G/5G parameters to explain NH performance, but lacked in a direct Wi-Fi comparison. This work bridges that gap by presenting a three-way comparative analysis of Wi-Fi, MNO macro, and NH performance---all utilizing the mid-band spectrum---within a large big-box store environment.
This study highlights the potential of indoor spectrum sharing for future mid-band allocations, such as the 3.1--3.45 GHz and 7.125--8.4 GHz bands currently being studied by the National Telecommunications and Information Administration (NTIA) for shared use~\cite{ntia2023strategy}.
Our contributions are as follows:

\noindent $\bullet$
\textbf{Comparison of coverage between NH, MNO macro, and Wi-Fi (\S\ref{sec:coverage_nh_mno}, \S\ref{sec:coverage_wifi}):} The NH network achieved better indoor coverage than the MNO macro deployment, requiring only six CBRS devices (CBSDs)---compared to the 65 access points (APs) required to achieve Wi-Fi coverage. Furthermore, a high median building loss of $26.6$~dB ensures fair coexistence between the NH network and outdoor incumbents.

\noindent $\bullet$
\textbf{Comparison of per-channel throughput performance between representative NH and MNO macro (\S\ref{subsec:pdsch_pusch_report}):} The NH network achieves median indoor normalized throughput gains of $1.44\times$ and $1.62\times$ in downlink (DL) and $4.33\times$ and $13\times$ in uplink (UL) compared to 4G and 5G macro deployments indoors, respectively. Moreover, our user equipment (UE) utilized lower uplink transmit (TX) power when utilizing neutral-host compared to MNO macro.

\noindent $\bullet$
\textbf{Comparison of user-experienced throughput performance between representative NH, MNO macro, and Wi-Fi (\S\ref{sec:app_report}):} The NH network achieves a median indoor aggregated PHY-layer DL throughput gain of $2.08\times$ over 5G macro deployments indoors, despite utilizing only $40$~MHz of aggregated bandwidth compared to $225$~MHz for 5G macro. Additionally, the NH deployment outperforms Wi-Fi in HTTP (Hypertext Transfer Protocol) DL throughput, achieving a $5.05\times$ improvement.

\section{Methodology and Tools Used} \label{sec:method_tools}


\begin{table}
    \centering
    \captionsetup{font=small}
    \caption{Summary of captured network performance parameters.}
    \label{tab:qp_params}
    \resizebox{.95\linewidth}{!}{ 
    \begin{tabular}{|C{8em}|L{25em}|}
        \hline
        \textbf{Parameter} & \hspace{8em} \textbf{Description} \\
        \hline\hline

        \multicolumn{2}{|c|}{\textbf{\textit{Qualipoc \& SigCap: General parameters}}} \\ \hline
        Latitude, Longitude & UE’s geographic coordinates \\ \hline
        \multicolumn{2}{|c|}{\textbf{\textit{Qualipoc: Radio report parameters}}} \\ \hline
        PCI & Physical Cell Identifier \\ \hline
        DL/UL ARFCN & Absolute Radio Frequency Channel Number, \ie center frequency. \\ \hline
        Bandwidth & Range of frequencies available for transmission {[}MHz{]} \\ \hline
        RSRP/RSRQ & Reference Signal Received Power and Reference Signal Received Quality values. For 5G, RSRP/RSRQ indicates measurements from the 5G Synchronization Signal (SS) block {[}dBm/dB{]} \\ \hline
        SCS & Subcarrier Spacing numerology; fixed at 15~kHz in 4G \\ \hline

        \multicolumn{2}{|c|}{\textbf{\textit{Qualipoc: Throughput \& power metrics}}} \\ \hline
        PDSCH/PUSCH Tput. & Throughput at Physical Downlink Shared Channel and Physical Uplink Shared Channel, \ie physical-layer throughput in downlink and uplink directions {[}Mbps{]} \\ \hline
        Normalized PDSCH/PUSCH throughput & A calculated metric which normalize physical layer throughput over the number of allocated resource blocks, subcarrier spacing, and MIMO layers {[}bit/s/Hz/stream{]} \cite{rochman2025comprehensive} \\ \hline
        App. DL/UL Tput. & Application layer throughput in downlink and uplink directions {[}Mbps{]} \\ \hline
        PUSCH TX power & Uplink TX power used by the UE {[}dBm{]} \\ \hline
        



        \multicolumn{2}{|c|}{\textbf{\textit{SigCap: Wi-Fi parameters}}} \\ \hline
        BSSID & Wi-Fi Basic Service Set Identifier that indicates unique identification of a Wi-Fi AP. \\ \hline
        Primary channel number & A number associated with a unique BSSID that identifies the frequency of the 20 MHz primary channel {[}MHz{]}. \\ \hline
        RSSI & Received Signal Strength Indicator calculated from a 20 MHz beacon signal {[}dBm{]}. \\ \hline
        TX power & Conducted power of the BSSID {[}dBm{]}. \\ \hline

    \end{tabular}
    }
    \vspace{-1em}
\end{table}

Our measurement campaign was conducted in a typical big-box retail store located in a suburban setting. The building is characterized by thick concrete walls and minimal windows. Within this store, users can be served by indoor neutral-host and Wi-Fi networks, or by outdoor-deployed cellular macro BSs.
For comparative measurements of the networks, we utilized a Samsung S22+ smartphone with connectivity to 2.4~GHz and 5~GHz Wi-Fi, as well as 4G and 5G cellular bands, including the CBRS band used by the neutral-host network. The device was equipped with two measurement tools: QualiPoc and SigCap.
The QualiPoc tool actively measure application-layer throughput by looping through a defined test sequence: a 5-seconds HTTP download from \textit{github.com} and a 5-seconds HTTP upload to \textit{httpbin.org}. Concurrently, cellular and Wi-Fi data were passively collected by QualiPoc and SigCap, respectively. 
QualiPoc collects detailed 4G \& 5G PHY-layer data by probing the modem chipset via the \textit{Qualcomm Diagnostic Mode} interface~\cite{qualipoc}.
Conversely, SigCap collects cellular and Wi-Fi data but it is limited to the information exposed by the Android API~\cite{sathya2020measurement}. Thus, we utilized SigCap exclusively for Wi-Fi analysis.
Table \ref{tab:qp_params} details QualiPoc and SigCap parameters used in our analysis.


The measurements were conducted by walking indoors and outdoors around the retail store, between Feb.~24 to Feb.~26, 2025, capturing $17{,}400$~m$^2$ ($187{,}292$~ft$^2$) of a large big-box retail store indoors and the adjacent $28{,}400$~m$^2$ ($305{,}695$~ft$^2$) parking lot outdoors. This store is served by three cellular MNOs labeled \att, \tmo, and \vzn. However, only \attandtmo offload their network to the neutral-host. Fig.~\ref{fig:map_coverage} shows the coverage footprint of our measurements as heatmaps of captured RSRP from MNO NH and MNO macros. We collected a total of 221,396 datapoints across QualiPoc radio reports, PDSCH (downlink) throughput, PUSCH (uplink) throughput, and application-layer (HTTP session) reports, as well as 570,803 Wi-Fi beacon datapoints from SigCap.

\section{Deployment Overview} \label{sec:deployment}


\begin{table}
\centering
\captionsetup{font=small}
\caption{4G and 5G bands information.}
\vspace{-0.5em}
\label{tab:nr-lte-bands}
\resizebox{0.85\linewidth}{!}{
\begin{tabular}{|C{.18\linewidth}|C{.1\linewidth}|C{.18\linewidth}|C{.12\linewidth}|C{.15\linewidth}|C{.1\linewidth}|}
\hline
\textbf{Operator-Band} & \textbf{Duplex Mode} & \textbf{DL Band Freq. (MHz)} & \textbf{SCS (kHz)} & \textbf{BW (MHz)} & \textbf{\#unique PCIs} \\ \hline \hline

\multicolumn{6}{|c|}{\textbf{NH band}} \\ \hline
\att b48 \tmo b48 & TDD & $3500$ & $15$ & $20$ & 12 \\ \hline

\multicolumn{6}{|c|}{\textbf{MNO macro bands}} \\ \hline

\att b2 & FDD & $1900$ & $15$ & $10$, $15$ & 5\\
\att b12 & FDD & $700$ & $15$ & $10$ & 5 \\
\att b14 & FDD & $700$ & $15$ & $10$ & 4 \\
\att b30 & FDD & $2300$ & $15$ & $10$ & 4 \\ 
\att b66 & FDD & $1700$ & $15$ & $10$ & 4 \\ \hline

\tmo b2 & FDD & $1900$ & $15$ & $5$ & 3 \\
\tmo b12 & FDD & $700$ & $15$ & $5$ & 2 \\
\tmo b66 & FDD & $1700$ & $15$ & $20$ & 3 \\ \hline

\vzn b2 & FDD & $1900$ & $15$ & $10$ & 3 \\
\vzn b5 & FDD & $850$ & $15$ & $10$ & 2 \\
\vzn b13 & FDD & $700$ & $15$ & $10$ & 4 \\
\vzn b66 & FDD & $1700$ & $15$ & $10$, $20$ & 4 \\ \hline

\rowcolor{light-gray}
\tmo n25 & FDD & $1900$ & $15$ & $10$, $15$ & 13 \\
\rowcolor{light-gray}
\tmo n41 & TDD & $2500$ & $30$ & $90$, $100$ & 34 \\
\rowcolor{light-gray}
\tmo n71 & FDD & $600$ & $15$ & $20$ & 21 \\ \hline

\rowcolor{light-gray}
\vzn n77 & TDD & $3700$ & $30$ & $100$ & 6 \\ \hline

\end{tabular}
}
\end{table}

\begin{table}
    \centering
    \captionsetup{font=small}
    \caption{Deployment parameters.}
    \vspace{-0.5em}
    \resizebox{0.8\linewidth}{!}{
    \begin{tabular}{|C{10.5em}|C{6.5em}|C{6.5em}|}
        \hline
        \textbf{Parameters} & \multicolumn{2}{c|}{\textbf{Values}} \\
        \hline
        \hline
        \multicolumn{3}{|c|}{\textbf{Neutral-host deployment}} \\
        \hline
        \# of deployed CBSD & \multicolumn{2}{c|}{6 CBSDs} \\
        \hline
        \# of PCI & \multicolumn{2}{c|}{2 per CBSD} \\
        \hline
        Nature of Antenna & \multicolumn{2}{c|}{Omnidirectional} \\
        \hline
        4G TDD Config & \multicolumn{2}{c|}{Config \#1: DL/UL (\# subframes): 4/4} \\
        \hline
        CBSD TX Power & \multicolumn{2}{c|}{$24$~dBm} \\
        \hline
        Antenna Gain & \multicolumn{2}{c|}{$3$~dBi} \\
        \hline
        Band Number \& \newline Channel Frequency & \multicolumn{2}{C{13em}|}{b48: \{$3560$, $3590$, $3620$, \newline $3650$, $3690$\}~MHz} \\
        \hline
        Bandwidth & \multicolumn{2}{c|}{$20$~MHz} \\
        \hline
        Channel aggregation & \multicolumn{2}{c|}{up to 2 channels ($40$~MHz)} \\
        \hline
        \hline
        \multicolumn{3}{|c|}{\textbf{Enterprise Wi-Fi deployment}} \\
        \hline
        \textbf{} & \textbf{2.4~GHz} & \textbf{5~GHz} \\
        \hline
        \# unique BSSIDs & 12 & 65 \\
        \hline
        Wi-Fi TX power & $14$~dBm & $10$--$19$~dBm \\
        \hline
        Bandwidth & \multicolumn{2}{c|}{$20$~MHz} \\
        \hline
        Wi-Fi standard & \multicolumn{2}{c|}{Wi-Fi 6 (IEEE~802.11ax)} \\
        \hline
    \end{tabular}
    }
    \label{table:exp_params}
    \vspace{-1em}
\end{table}

\textit{\textbf{Neutral-Host \& MNO Macro:}} Table~\ref{tab:nr-lte-bands} provides a summary of the 4G and 5G bands observed during the measurement campaign, where 4G bands are denoted with the prefix~\textit{`b'} and 5G bands with the prefix~\textit{`n'}.
Among these deployments, \attandtmo offloads their indoor users to the neutral-host services in the CBRS b48 band. The NH deployment within the retail store consists of six indoor CBSDs, each configured with two Physical Cell Identifiers (PCIs), resulting in a total of 12 active PCIs. These CBSDs are deployed on ceilings (approximately 4--5~m above ground level) and employ omnidirectional antennas with a TX power and antenna gain of $24$~dBm and $3$~dBi to uniformly distribute coverage within the indoor environment. The deployment utilizes five unique $20$~MHz channels within the b48 band, enabling channel reuse and segmentation across spatial zones, and also supports aggregation of up to two channels (totaling $40$~MHz) for higher capacity. More details on the NH deployment are provided in Table~\ref{table:exp_params}.

In the cellular macro deployment, all three MNOs have deployed 4G across both low-band ($<$1~GHz) and mid-band (1--6~GHz) frequencies. 5G deployments were observed for both \tmo and \vzn, with higher capacity observed in its mid-band deployments (\tmo's n41 and \vzn's n77) as shown by their higher sub-carrier spacing (SCS) and wider bandwidth. Notably, \tmo demonstrates a dense and diverse 5G deployment with high number of unique PCI across low- and mid-band frequencies.


\textit{\textbf{Enterprise Wi-Fi:}} To focus our analysis on the retail store's enterprise Wi-Fi network, we filtered beacon data for its specific Service Set Identifier (SSID).
Our measurements identified 65 unique 5~GHz and 12 unique 2.4~GHz Basic Service Set Identifier (BSSIDs), indicating a dense 5~GHz deployment designed for comprehensive indoor coverage. These APs are deployed at on ceilings at the same heights as the CBSDs. They operate with a fixed 20~MHz channel width, suggesting the higher bandwidth allowed in IEEE~802.11ax ($40$--$160$~MHz) were likely not configured, potentially to avoid co-channel interference or to prioritize interoperability with legacy clients. Beacon TX power data further reveals that the 2.4~GHz radios operate at a fixed TX power of $14$~dBm, and a range from $10$~dBm to $19$~dBm for the 5~GHz radios, indicative of adaptive power control likely employed to optimize coverage and mitigate co-channel interference.




\section{Measurement Results and Analyses}\label{sec:results}

\subsection{Coverage Comparison of NH and MNO macro}\label{sec:coverage_nh_mno}

For an overall comparison of coverage between the neutral-host and MNO macro deployments, we refer to the RSRP heatmaps in Fig.~\ref{fig:map_coverage}. Specifically, Fig.~\ref{fig:map_mno_nh} illustrates the RSRP heatmap from the CBRS band (b48) utilized by the NH deployment under both \attandtmo. Each square of the heatmap represents the maximum RSRP over all datapoints within the $6\times 6$~m$^2$ bin. Figs.~\ref{fig:map_mno_lte} and \ref{fig:map_mno_nr} summarizes the coverage of all MNO macros, by combining maximum RSRP captured on all 4G and 5G channels, respectively. We observe higher indoor RSRP from neutral-host compared to macro deployments, and vice versa for outdoor RSRP---an expected outcome given the indoor deployment of neutral-host and the outdoor deployment of MNO macro.



\begin{figure}
\centering
\includegraphics[width=.95\linewidth]{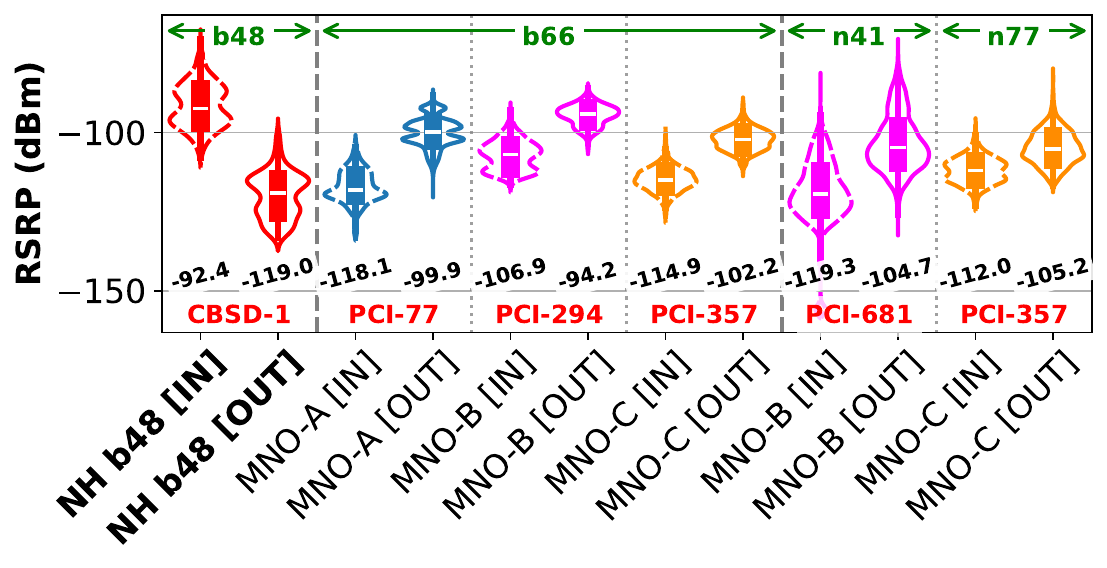}
\caption{Coverage statistics for representative CBSD/PCI.}
\label{fig:radio_pci_rsrp}
\vspace{-1.5em}
\end{figure}

\begin{figure*}
    \begin{subfigure}{.45\textwidth}
    \includegraphics[width=\linewidth]{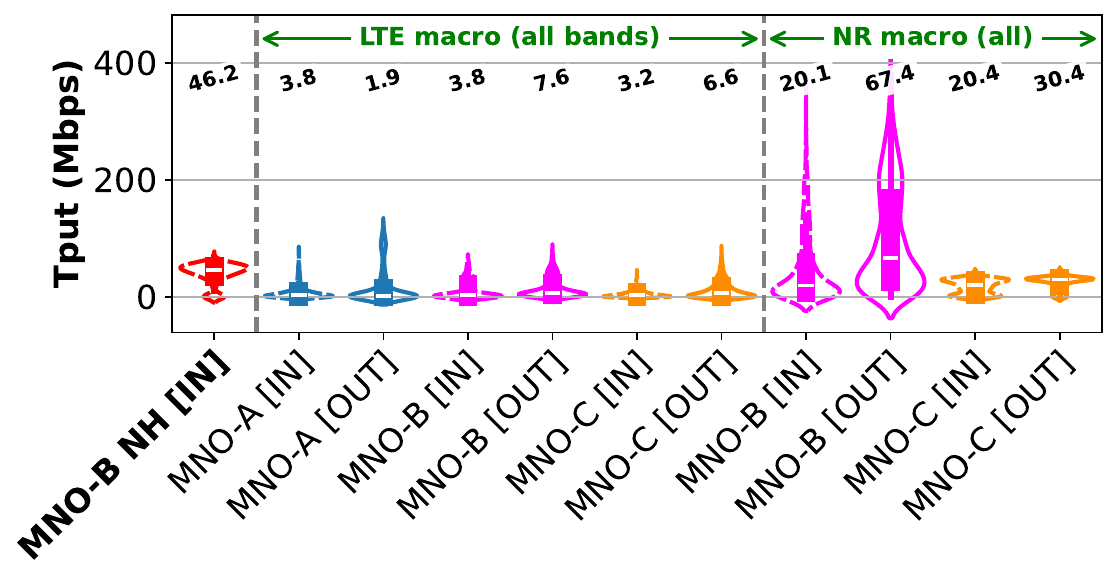}
    \vspace{-2em}
    \caption{PDSCH throughput.}
    \label{fig:pdsch_all_tput}
    \end{subfigure}
    \hfill
    \begin{subfigure}{.45\textwidth}
    \includegraphics[width=\linewidth]{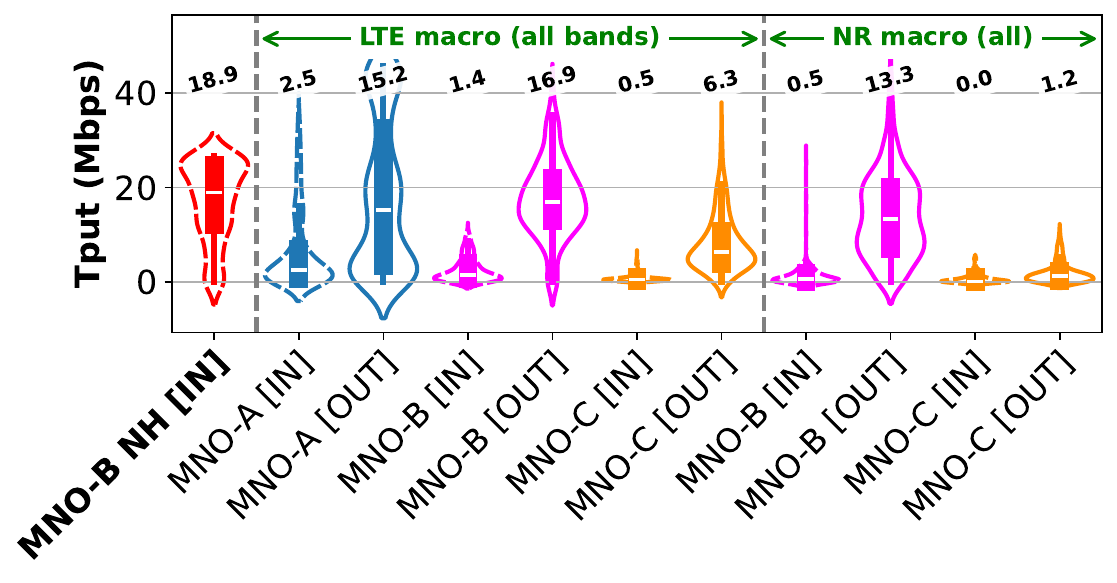}
    \vspace{-2em}
    \caption{PUSCH throughput.}
    \label{fig:pusch_all_tput}
    \end{subfigure}
    \caption{PHY-layer throughput comparison across NH and MNO macro deployments.}
    \label{fig:phy_performance}
    \vspace{-.5em}
\end{figure*}

\begin{figure*}
    \begin{subfigure}{.3\textwidth}
    \includegraphics[width=\linewidth]{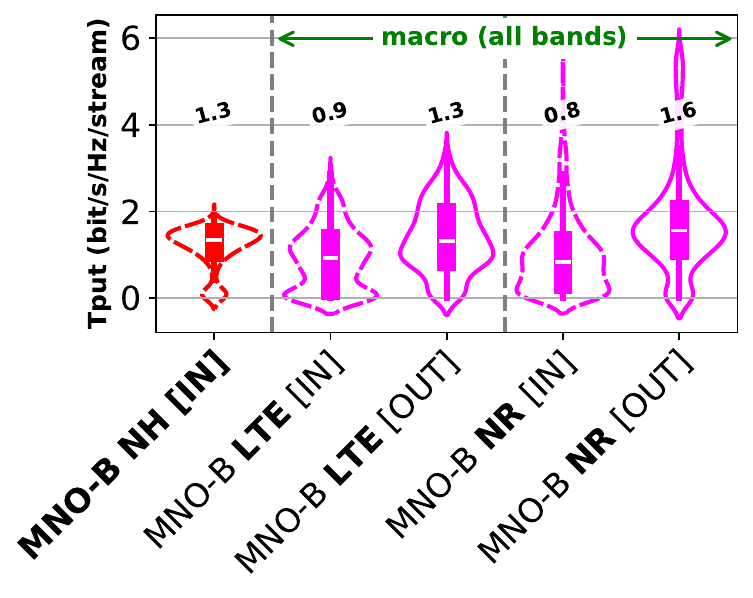}
    \vspace{-1.5em}
    \caption{Normalized PDSCH throughput.}
    \label{fig:pdsch_select_normtput}
    \end{subfigure}
    \hfill
    \begin{subfigure}{.3\textwidth}
    \includegraphics[width=\linewidth]{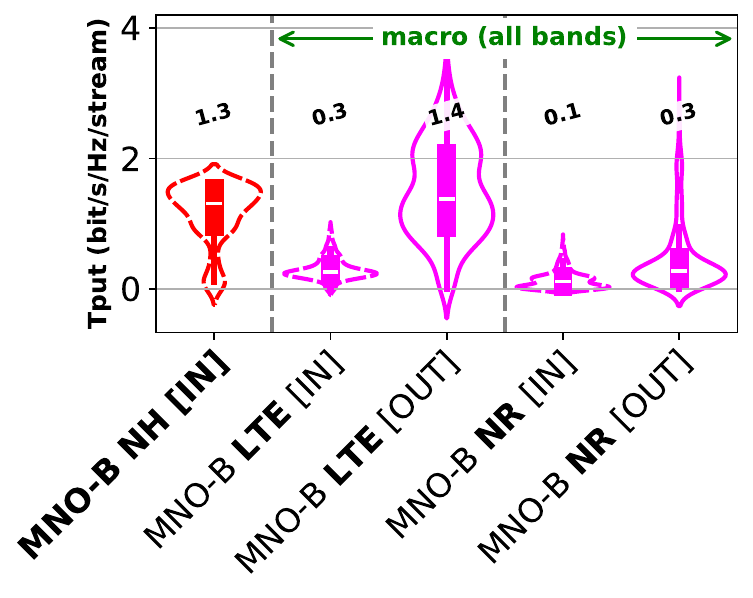}
    \vspace{-1.5em}
    \caption{Normalized PUSCH throughput.}
    \label{fig:pusch_select_normtput}
    \end{subfigure}
    \hfill
    \begin{subfigure}{.3\textwidth}
    \includegraphics[width=\linewidth]{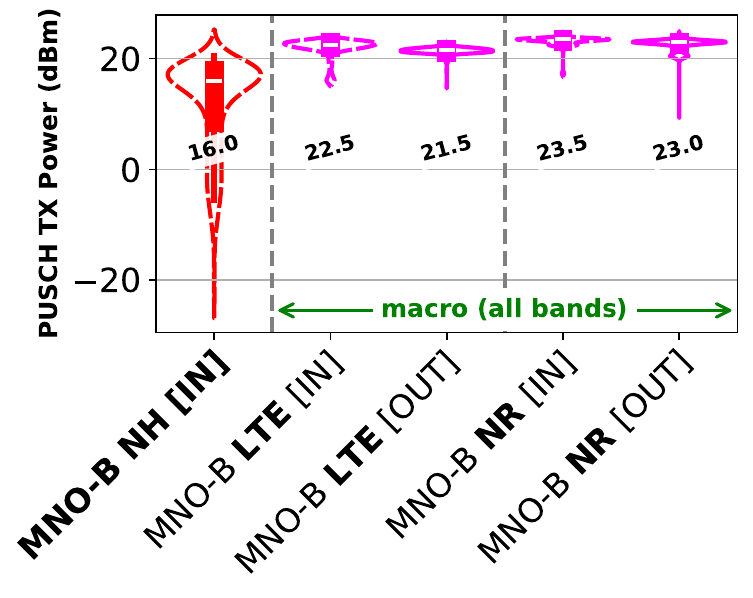}
    \vspace{-1.5em}
    \caption{PUSCH TX power distribution.}
    \label{fig:pusch_select_txpow}
    \end{subfigure}
    \caption{PHY-layer performance analysis of \tmo's NH and macro deployments.}
    \label{fig:select_phy_performance}
    \vspace{-1em}
\end{figure*}

Fig.~\ref{fig:radio_pci_rsrp} further illustrates the RSRP distributions for representative PCIs from the NH and MNO macro deployments, grouped by band. These representative PCIs were selected based on having the highest number of measurement points, serving as a proxy for primary coverage cells within the region of analysis. 
The representative CBSD (``CBSD-1'')
exhibits a significant building loss with a median indoor-to-outdoor RSRP difference of $26.6$~dB.
Further, the median RSRP values of the representative MNO macro deployments are high outdoors but degrades substantially indoors.
This underscore the limited indoor coverage of outdoor macro deployments even with a potentially high TX power. In contrast, CBSD-1 achieves a notably higher median indoor RSRP of $-92.4$~dBm while operating at a modest TX power of $24$~dBm.
This highlights the NH deployment’s ability to ensure strong indoor coverage while avoiding outdoor interference.


\subsection{PHY-layer Performance Comparison of NH and MNO macro}\label{subsec:pdsch_pusch_report}



Because NH traffic routed to MNO core networks is often treated as ``roaming'', users may face operator-specific throttling. This was confirmed by the NH infrastructure provider and is evident in our data, with a \tmo-subscribed device achieving $21$~Mbps higher median PDSCH throughput indoors than an \att-subscribed one.
To ensure a fair and representative evaluation of the NH deployment’s capabilities, we therefore focus exclusively on the performance analysis of \tmo's NH deployment throughout this study.


For a comprehensive view of the MNO macro deployments, Fig. \ref{fig:phy_performance} shows the per-channel PHY-layer throughput results for all MNOs and bands, with Figs. \ref{fig:pdsch_all_tput} and \ref{fig:pusch_all_tput} showing the PHY-layer DL and UL performance, respectively. Rather than focusing on representative PCIs---which may introduce site-specific biases---we present an overall comparison by grouping the measurements according to radio access technology (4G LTE or 5G NR) and location type (indoor/outdoor). Given \vzn’s low indoor DL and UL performance and its lack of traffic offload to the NH infrastructure, it is clear that \vzn could benefit from the superior performance provided by neutral-hosts.
In particular, \tmo’s macro 5G network delivers superior outdoor performance in the downlink direction. 
For UL throughput, the NH network's indoor performance is comparable to the outdoor performance of \att in 4G as well as \tmo in 4G and 5G.
Since our NH analysis uses data from the representative \tmo network, we focus exclusively on \tmo’s NH and macro deployments for the rest of our evaluation to maintain consistency and fairness.

\begin{figure*}
\begin{subfigure}{.28\textwidth}
\includegraphics[width=\linewidth]{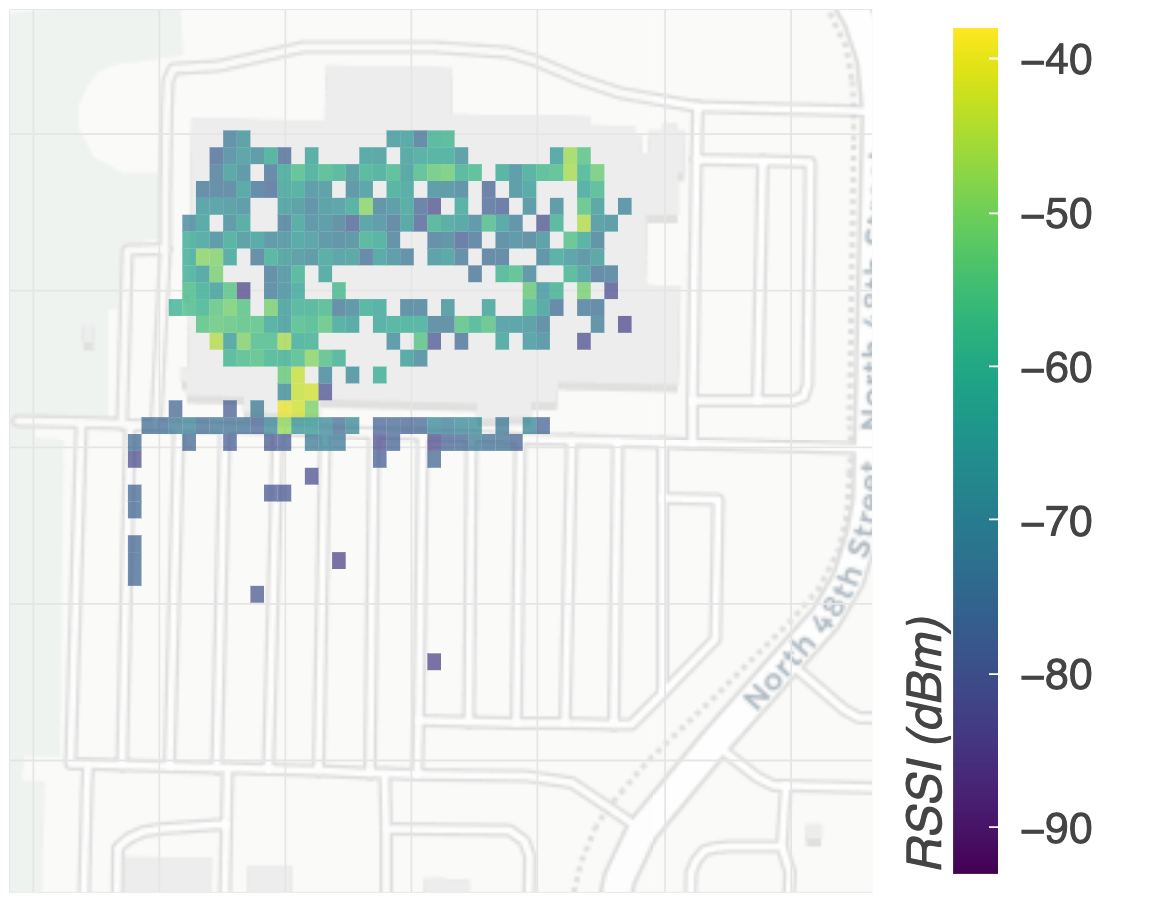}
\vspace{-1.5em}
\caption{Wi-Fi 2.4~GHz RSSI map}
\label{fig:map_wifi_2ghz_rssi}
\end{subfigure}
\hfill
\begin{subfigure}{.28\textwidth}
\includegraphics[width=\linewidth]{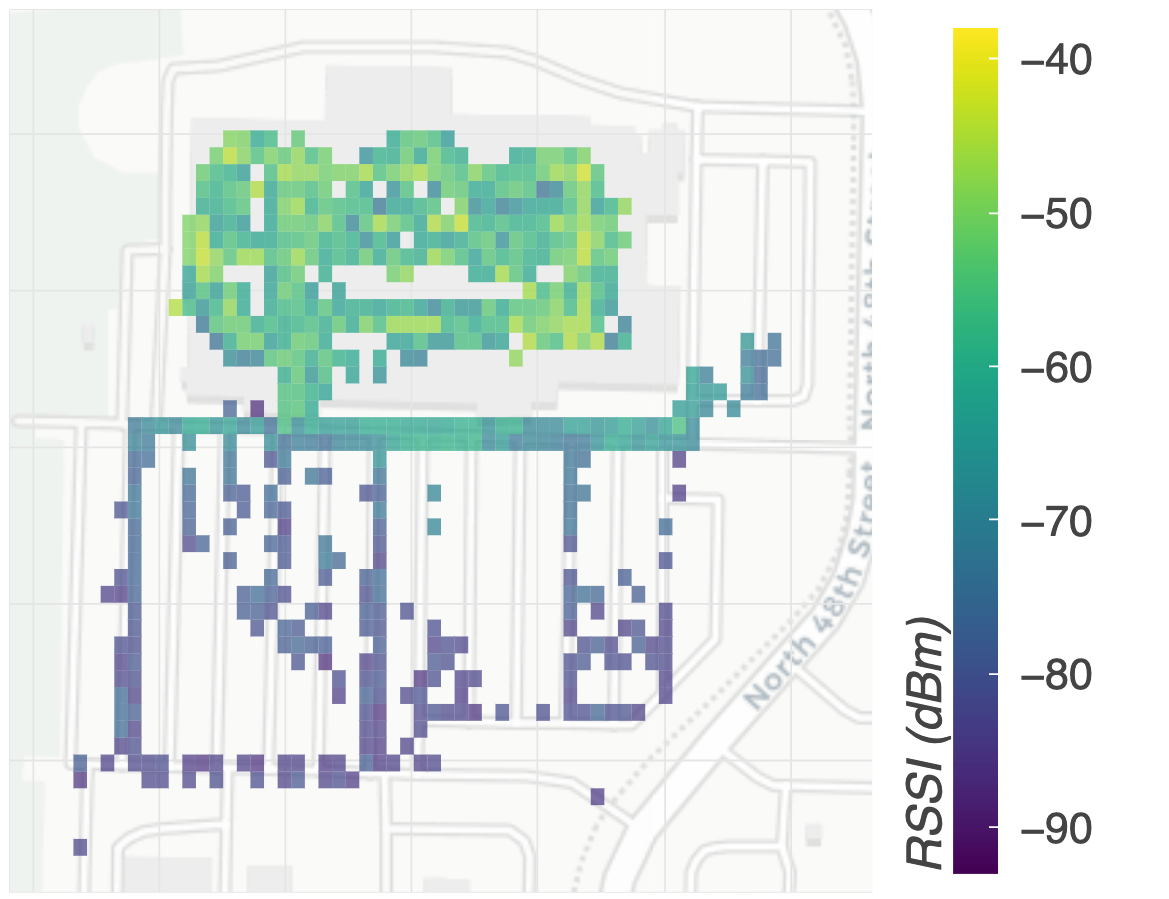}
\vspace{-1.5em}
\caption{Wi-Fi 5~GHz RSSI map}
\label{fig:map_wifi_5ghz_rssi}
\end{subfigure}
\hfill
\begin{subfigure}{.38\textwidth}
\includegraphics[width=\linewidth]{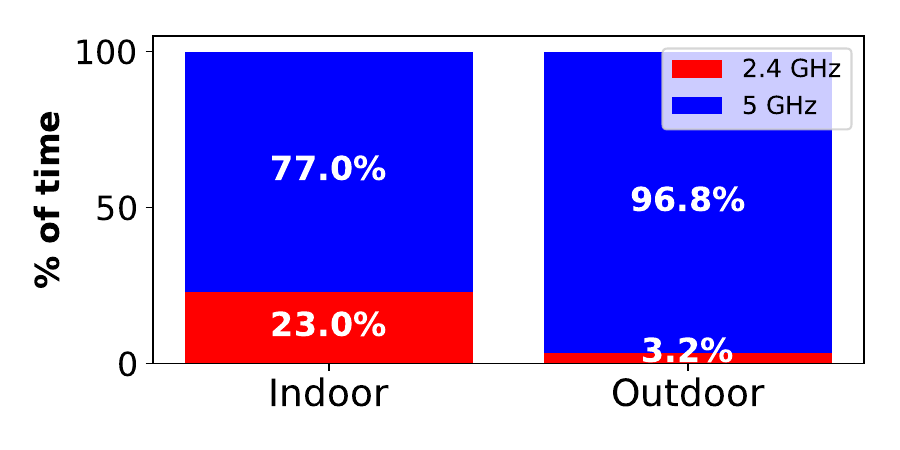}
\vspace{-1.5em}
\caption{Connected Wi-Fi band usage.}
\label{fig:wifi_conn_time}
\end{subfigure}
\caption{Wi-Fi coverage maps and time-based band usage statistics.}
\label{fig:wifi_rssi_map_cdf}
\vspace{-1.5em}
\end{figure*}

To enable a deeper analysis of radio link performance under fixed resource conditions, we normalize the PHY-layer throughput over the number of allocated resource blocks, subcarrier spacing, and MIMO layers: effectively capturing the channel spectral efficiency per spatial stream. We employ this metric---expressed in bit/s/Hz/stream---for comparison purposes, with its detailed formulation available in \cite{rochman2025comprehensive}.
The resulting distribution of normalized PDSCH and PUSCH throughput are shown in Figs.~\ref{fig:pdsch_select_normtput} and \ref{fig:pusch_select_normtput}, respectively. Looking at the normalized DL results in Fig.~\ref{fig:pdsch_select_normtput}, our \tmo-subscribed indoor user achieves a median throughput of $1.3$~bit/s/Hz/stream when served by the NH network, which represents gains of $1.44\times$ and $1.62\times$ over \tmo's own 4G and 5G macro deployments, respectively. Fig.~\ref{fig:pusch_select_normtput} shows a median normalized UL throughput of $1.3$~bit/s/Hz/stream for neutral-host, which is similar to its DL counterpart and consistent with its balanced 4G TDD configuration (4 subframes each for DL and UL). Notably, indoor UL performance gain over 4G and 5G macro deployments is even higher than that observed in the DL, with median gains of $4.33\times$ for 4G and $13\times$ for 5G.
These results underscores the effectiveness of the indoor NH system, which---despite consisting of only six low-power CBSDs---matches the spectral efficiency of resource-intensive outdoor macro deployments. Notably,
the NH deployment achieves comparable normalized DL performance while relying on a significantly smaller infrastructure footprint and lower TX power.

To complement the PHY-layer performance analysis, Fig.~\ref{fig:pusch_select_txpow} reveals that PUSCH TX power required by the UE is significantly lower when utilizing the NH network. In contrast, macro networks exhibits higher UE TX power usage, suggesting a potential advantage of neutral-host model in terms of UE's energy efficiency. 


While the QualiPoc tool is capable of capturing a wide range of PHY-layer parameters, we omit detailed analyses here, as a comprehensive PHY-layer study of the NH deployment has already been presented in our prior work~\cite{rochman2025neutralhosts} and is not repeated here. These findings collectively suggest that NH deployments offer not only competitive DL/UL performance but also reduced UE power demands, contributing to a more energy-efficient user-experience.








\subsection{Coverage Analysis of Enterprise Wi-Fi}\label{sec:coverage_wifi}


We assess the enterprise Wi-Fi coverage quality by analyzing the RSSI heatmaps in Figs. \ref{fig:map_wifi_2ghz_rssi} and \ref{fig:map_wifi_5ghz_rssi}.
Additionally, Fig.~\ref{fig:wifi_conn_time} presents the percentage of time UE was connected to the 2.4~GHz and 5~GHz bands, both indoors and outdoors. These results show that the 5~GHz band has more coverage (consistent with its higher AP deployment density), and that the UE prefers the 5~GHz band 77\% of the time indoors. Based on this, we use the 5~GHz band as the representative Wi-Fi band in all future analyses to ensure a fair comparison.

\begin{figure}
\centering
\includegraphics[width=0.49\textwidth]{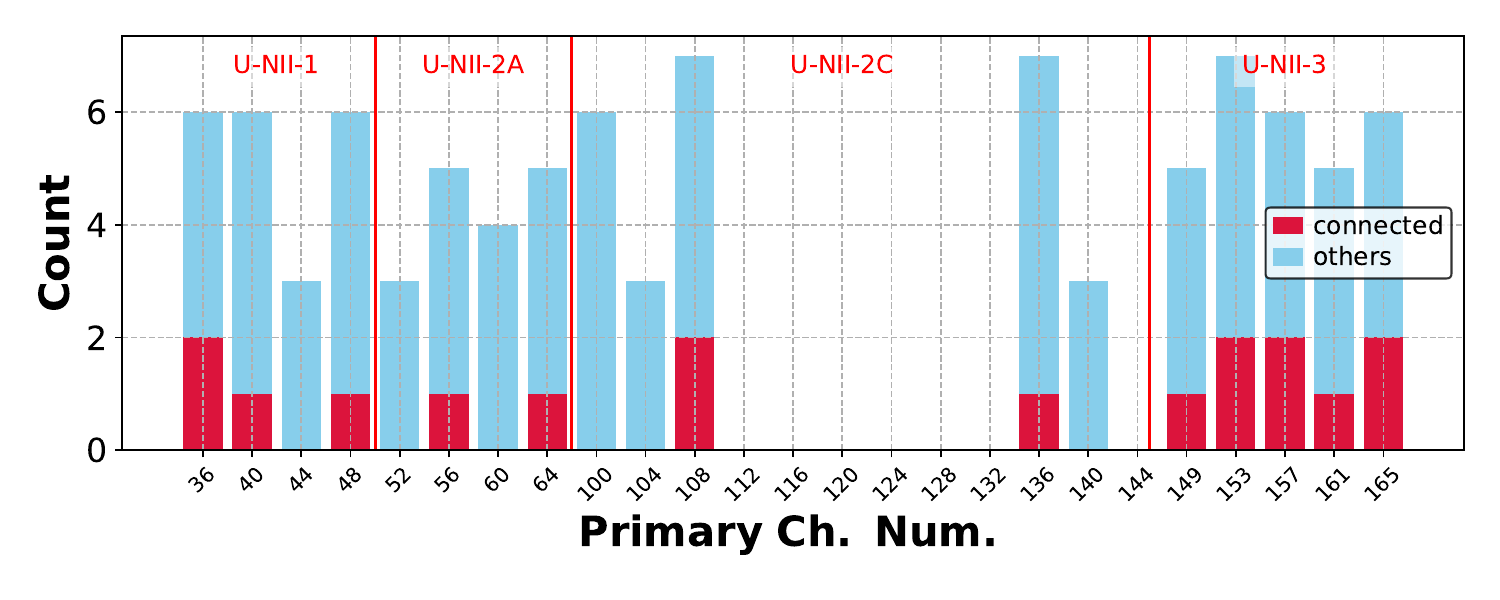}
\caption{Number of unique BSSIDs across 5~GHz channels}
\label{fig:wifi_channel_count}
\vspace{-1em}
\end{figure}

\begin{figure*}
\begin{subfigure}{.3\textwidth}
\includegraphics[width=\linewidth]{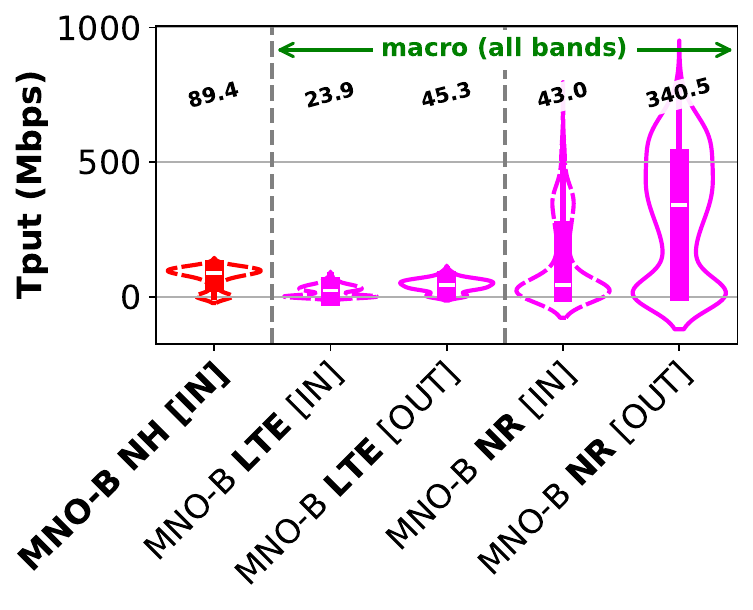}
\vspace{-1.5em}
\caption{Aggregated PDSCH throughput.}
\label{fig:pdsch_agg_select_tput}
\end{subfigure}
\hfill
\begin{subfigure}{.3\textwidth}
\includegraphics[width=\linewidth]{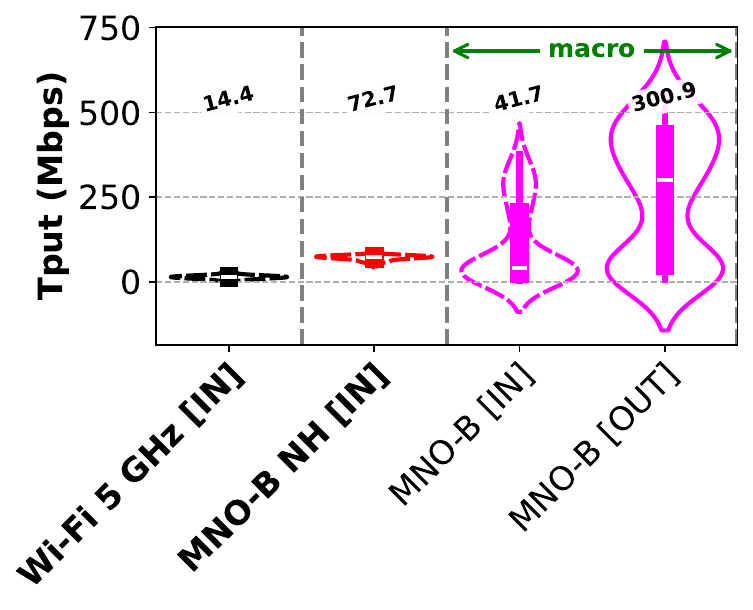}
\vspace{-1.5em}
\caption{HTTP DL throughput.}
\label{fig:app_select_dl_tput_violin}
\end{subfigure}
\hfill
\begin{subfigure}{.3\textwidth}
\includegraphics[width=\linewidth]{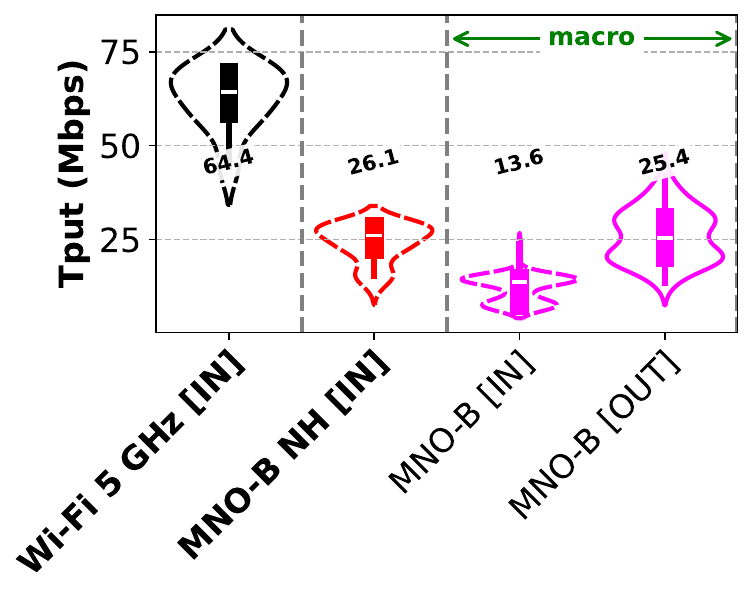}
\vspace{-1.5em}
\caption{HTTP UL throughput.}
\label{fig:app_select_ul_tput_violin}
\end{subfigure}
\caption{User-experienced throughput statistics.}
\label{fig:user_throughput}
\vspace{-1.5em}
\end{figure*}

To further understand 5~GHz Wi-Fi spectrum utilization, Fig.~\ref{fig:wifi_channel_count} presents the number of unique BSSIDs both scanned and connected to across the available 5~GHz channels. The distribution is generally uniform across most channels, indicating a balanced channel assignment. However, channels within the U-NII-2C sub-band are notably underutilized, with several channels not being used at all. This may be attributed to dynamic frequency selection (DFS) restrictions: U-NII-1 and U-NII-3 bands might be more preferable to maintain interoperability with Wi-Fi clients that does not support DFS.
The extensive 5~GHz coverage---achieved with 65 APs---contrasts sharply with the six CBSDs utilized by the NH network, highlighting the NH deployment’s efficiency in delivering robust indoor coverage with significantly fewer infrastructure resources.

\subsection{User-Experience Throughput Comparison Between NH, MNO macro, and Wi-Fi}\label{sec:app_report}

While \S\ref{subsec:pdsch_pusch_report} provided a per-channel PHY-layer throughput analysis, it is more relevant from a user-experience perspective to consider the aggregated throughput across all connected bands, as shown in Fig.~\ref{fig:pdsch_agg_select_tput}. \tmo’s 5G macro network (operating in 5G standalone mode) achieves a substantially higher median aggregated PDSCH throughput, benefiting from the aggregation of up to four 5G channels (totaling $225$~MHz). In comparison, the NH deployment aggregates only two 4G channels ($20$~MHz each, for a total of $40$~MHz), while \tmo’s 4G macro uses up to three 4G channels (totaling $30$~MHz). Interestingly, despite the higher bandwidth available to the \tmo macro 5G, the indoor performance of neutral-host outperforms the 5G macro $2.08\times$ times. This indicates that, even though NH operates on 4G with limited bandwidth, its indoor-based deployment yields superior performance compared to outdoor-deployed 5G---demonstrating that deployment location can be just as crucial as spectrum resources in determining system performance.

To incorporate Wi-Fi into the end-user performance evaluation, we compare DL and UL application-layer throughput across neutral-host, MNO macro, and enterprise Wi-Fi deployments. The throughput was measured using controlled HTTP GET and PUT requests, corresponding to DL and UL traffic, respectively. For each session, application-layer throughput was calculated by dividing the total amount of data transferred by the session duration.
Thus, the measured HTTP-layer throughput represents aggregate performance across active radio channels, encompassing scenarios such as 4G-only (aggregating solely 4G channels), 5G Standalone (solely 5G channels), or 5G Non-Standalone (combinations of 4G and 5G channels).
Conversely, the application-layer throughput under Wi-Fi networks remains constrained to the $20$~MHz channel used. The results are summarized in Figs.~\ref{fig:app_select_dl_tput_violin} and \ref{fig:app_select_ul_tput_violin}, which respectively show the distribution of DL and UL application-layer throughput across technologies and environment.

Fig.~\ref{fig:app_select_dl_tput_violin} shows the indoor DL performance of \tmo's NH outperforms enterprise Wi-Fi with a $5.05\times$ improvement.
Further comparisons with \tmo's macro deployments corroborate the aggregated PHY-layer PDSCH throughput results presented in Fig.~\ref{fig:pdsch_agg_select_tput}.
In the UL, Fig.~\ref{fig:app_select_ul_tput_violin} shows that Wi-Fi achieves a notably high median UL throughput of $64.4$~Mbps indoors---surpassing its own DL performance. This asymmetry is likely due to the difference in target host utilized in the HTTP GET and PUT request.
Excluding Wi-Fi, the NH network again shows strong UL performance comparable to outdoor macro performance and exceeding indoor macro performance by $1.92\times$.

Taken together, these results demonstrate that the NH network provides competitive user-experience performance when compared to indoor Wi-Fi and indoor MNO macro deployments, while also remaining competitive with macro-cellular performance in outdoor scenarios---when evaluated critically for bandwidth resource utilization. Our evaluation considered both aggregated PHY-layer PDSCH throughput and application-layer HTTP throughput as metrics for assessing user-experienced performance, ensuring a comprehensive view of network effectiveness.







\section{Conclusions and Future Work}\label{sec:conclusions}

This study presents a comprehensive in-situ evaluation of mid-band deployments---enterprise Wi-Fi, MNO macro, and neutral-host networks---at a large big-box store, focusing on coverage, PHY-layer and application-layer throughput. Our measurements reveal that while Wi-Fi requires 65 APs to achieve extensive indoor coverage, the NH deployment provides comparable coverage with just six CBSDs and offers superior indoor coverage compared to MNO macro deployments. Analyzing PHY-layer performance, the NH network consistently outperforms the MNO macro indoors in both downlink and uplink, for both per-channel and aggregated throughput. Application-layer HTTP downlink throughput analysis further reinforces that the NH network provides a superior user experience compared to Wi-Fi and indoor MNO macro deployments, making it a strong candidate for future mid-band indoor connectivity solutions.

\bibliographystyle{IEEEtran}
\bibliography{IEEEabrv,main}

\begin{thebibliography}{10}
\providecommand{\url}[1]{#1}
\csname url@samestyle\endcsname
\providecommand{\newblock}{\relax}
\providecommand{\bibinfo}[2]{#2}
\providecommand{\BIBentrySTDinterwordspacing}{\spaceskip=0pt\relax}
\providecommand{\BIBentryALTinterwordstretchfactor}{4}
\providecommand{\BIBentryALTinterwordspacing}{\spaceskip=\fontdimen2\font plus
\BIBentryALTinterwordstretchfactor\fontdimen3\font minus \fontdimen4\font\relax}
\providecommand{\BIBforeignlanguage}[2]{{%
\expandafter\ifx\csname l@#1\endcsname\relax
\typeout{** WARNING: IEEEtran.bst: No hyphenation pattern has been}%
\typeout{** loaded for the language `#1'. Using the pattern for}%
\typeout{** the default language instead.}%
\else
\language=\csname l@#1\endcsname
\fi
#2}}
\providecommand{\BIBdecl}{\relax}
\BIBdecl

\bibitem{ericsson2023mobility}
{Ericsson}, ``{Ericsson Mobility Report},'' Retrieved from \url{https://www.ericsson.com/4ae12c/assets/local/reports-papers/mobility-report/documents/2023/ericsson-mobility-report-november-2023.pdf}, Nov. 2023, accessed: May 2025.

\bibitem{calabrese2024frontier}
M.~Calabrese and J.~Dine, ``The next frontier in spectrum policy: Indoor-only sharing of federal bands,'' Retrieved from \url{https://d1y8sb8igg2f8e.cloudfront.net/documents/The_Next_Frontier_in_Spectrum_Policy_2024-11-25_141303_RP4Nq3g.pdf}, Nov. 2024, accessed: May 2025.

\bibitem{dogan2023evaluating}
S.~Dogan-Tusha, M.~I. Rochman, A.~Tusha, H.~Nasiri, J.~Helzerman, and M.~Ghosh, ``Evaluating the interference potential in 6 {GHz}: An extensive measurement campaign of a dense indoor {Wi-Fi} {6E} network,'' in \emph{Proceedings of the 17th ACM Workshop on Wireless Network Testbeds, Experimental evaluation \& Characterization}, 2023, pp. 56--63.

\bibitem{dogan2023indoor}
S.~Dogan-Tusha, A.~Tusha, H.~Nasiri, M.~I. Rochman, and M.~Ghosh, ``Indoor and outdoor measurement campaign for unlicensed 6 {GHz} operation with {Wi-Fi 6E},'' in \emph{2023 26th International Symposium on Wireless Personal Multimedia Communications (WPMC)}.\hskip 1em plus 0.5em minus 0.4em\relax IEEE, 2023, pp. 1--6.

\bibitem{tusha2025comprehensive}
A.~Tusha, S.~Dogan-Tusha, J.~R. Palathinkal, H.~Nasiri, M.~I. Rochman, P.~McGuire, and M.~Ghosh, ``A comprehensive analysis of secondary coexistence in a real-world {CBRS} deployment,'' \emph{IEEE Transactions on Cognitive Communications and Networking}, 2025.

\bibitem{bajracharya2022neutral}
R.~Bajracharya, R.~Shrestha, H.~Jung, and H.~Shin, ``Neutral host technology: The future of mobile network operators,'' \emph{IEEE Access}, vol.~10, pp. 99\,221--99\,234, 2022.

\bibitem{sathya2023nh}
V.~Sathya, M.~Shah, and M.~Yavuz, ``Neutral host deployment-a measurement study on routing and mobility management between public and private network,'' in \emph{2023 IEEE Future Networks World Forum (FNWF)}.\hskip 1em plus 0.5em minus 0.4em\relax IEEE, 2023, pp. 1--8.

\bibitem{sathya2023warehouse}
V.~Sathya, L.~Zhang, M.~Goyal, and M.~Yavuz, ``Warehouse deployment: A comparative measurement study of commercial {Wi-Fi} and {CBRS} systems,'' in \emph{2023 International Conference on Computing, Networking and Communications (ICNC)}.\hskip 1em plus 0.5em minus 0.4em\relax IEEE, 2023, pp. 242--248.

\bibitem{rochman2025neutralhosts}
\BIBentryALTinterwordspacing
M.~I. Rochman, J.~R. Palathinkal, V.~Sathya, M.~Yavuz, and M.~Ghosh, ``Neutral-hosts in the shared mid-bands: Addressing indoor cellular performance,'' 2025. [Online]. Available: \url{https://arxiv.org/abs/2505.18360}
\BIBentrySTDinterwordspacing

\bibitem{ntia2023strategy}
{National Telecommunications and Information Administration}, ``{The National Spectrum Strategy},'' Retrieved from \url{https://www.ntia.gov/sites/default/files/publications/national_spectrum_strategy_final.pdf}, 2023, accessed: Dec. 2024.

\bibitem{rochman2025comprehensive}
M.~I. Rochman, W.~Ye, Z.-L. Zhang, and M.~Ghosh, ``A comprehensive real-world evaluation of {5G} improvements over {4G} in low-and mid-bands,'' \emph{IEEE Transactions on Cognitive Communications and Networking}, 2025.

\bibitem{qualipoc}
{Rohde \& Schwarz}, ``{QualiPoc Android},'' Retrieved from \url{https://www.rohde-schwarz.com/us/products/test-and-measurement/network-data-collection/qualipoc-android\_63493-55430.html}, accessed: Dec. 2024.

\bibitem{sathya2020measurement}
V.~Sathya, M.~I. Rochman, and M.~Ghosh, ``Measurement-based coexistence studies of {LAA} \& {Wi-Fi} deployments in chicago,'' \emph{IEEE Wireless Communications}, vol.~28, no.~1, pp. 136--143, 2020.

\end{thebibliography}

\end{document}